\documentclass[cpp,fleqn]{w-art}

\usepackage{amsmath}
\usepackage{amssymb}
\usepackage{times}
\usepackage{w-thm}
\usepackage{wrapfig}
\usepackage[]{graphicx}

\def\respace{\mathpalette\mathllapinternal}
\def\mathllapinternal#1#2{%
\makebox[10em][l]{$\mathsurround=0pt#1{#2}$}}
\DeclareMathOperator{\erfc}{erfc}
\begin{document}

\keywords{integral equations, static structure factor, two-component plasma, non-equilibrium}
\subjclass{02.30.Rz, 52.27.Gr, 52.27.Aj, 05.70.Ln}
\title[HNC calculations for TCP]{Hypernetted chain calculations for two-component plasmas}
\author[Schwarz~\textit{et~al.}]{V. Schwarz\footnote{Corresponding author:~\textsf{volker.schwarz@uni-rostock.de}}\inst{1} \and Th. Bornath\inst{1} \and W. D. Kraeft\inst{1,2} \and S. Glenzer\inst{3} \and A. H\"oll\inst{1}\and R. Redmer\inst{1}} %

\DOIsuffix{theDOIsuffix} %%

\Volume{42} \Issue{1} \Month{01} \Year{2003} %%
\pagespan{1}{} %%

%\Receiveddate{15 November 2003} 
%\Reviseddate{30 November 2003} %
%\Accepteddate{2 December 2003} 
%\Dateposted{3 December 2003} %%

\address[\inst{1}]{Institut f\"ur Physik, Universit\"at Rostock, 18051 Rostock, Germany} %%
\address[\inst{2}]{Institut f\"ur Physik, Ernst-Moritz-Arndt-Universit\"at, 17487 Greifswald, Germany} %%
\address[\inst{3}]{L-399, Lawrence Livermore National Laboratory, University of California, P.O. Box 808, Livermore, CA 94551, USA} %%

\begin{abstract}
We have performed HNC calculations for dense beryllium plasma as studied experimentally using x-ray Thomson scattering, recently. We treated non-equilibrium situations with different electron and ion temperatures which are relevant in pump-probe experiments on ultra-short time scales. To consider quantum effects adequately, we used effective pair potentials to describe the interactions. Results are compared with classical as well as quantum corrected Debye model calculations.
\end{abstract}
\maketitle
\section{\label{Introduction}Introduction}
The dynamic structure factor $S\left(k,\omega\right)$ is the spectral function of the density fluctuations. It determines fundamental properties of the plasma such as pair distribution functions, equation of state data, and transport coefficients~\cite{Ichimaru86}. Simultaneously, it gives direct access to the frequency spectrum of amplitude-modulated electromagnetic waves scattered off the plasma and, therefore, is a key quantity for plasma diagnostics. We are especially interested in strongly coupled plasmas with densities typical of solid state and temperatures of several eV which are also known as \textit{warm dense matter} (WDM). Intense x-ray radiation sources are needed to perform scattering experiments in WDM targets. Free electron laser facilities like FLASH at DESY in Hamburg~\cite{Tschentscher05} will be available to deliver such radiation. Alternatively, pump-probe experiments are also planned at GSI Darmstadt using the future FAIR facility \cite{Tahir06}.\par
Pioneering pump-probe experiments on warm dense beryllium have been performed at the Laboratory for Laser Energetics with the Omega laser in Rochester. They were successful in spectrally resolving the non-collective (particle) scattering characteristics of beryllium~\cite{Glenzer03} and carbon~\cite{Gregori06a}. Recently, it has been shown~\cite{Glenzer06} that important plasma parameters such as electron density and temperature can be inferred from the collective scattering feature (plasmons) of the measured dynamic structure factor.\par
Therefore, a detailed analysis of the dynamic structure factor $S\left(k,\omega\right)$ in WDM states is needed in order to account for strong correlations and scattering processes between the particles which determine the damping of excitations processes~\cite{Gregori03,Redmer05}. Time resolved measurements on ultra-short time scales would offer the possibility to study non-equilibrium states, e.g. different electron and ion temperature and the relaxation into thermodynamic equilibrium, see~\cite{Gregori06b,Gericke02a,Gericke02b,Cauble83}.\par
We present an approach describing a plasma consisting of electrons and ions having different temperatures. In order to calculate the static structure factor $S\left(k\right)$ for dense beryllium plasma we solve an Ornstein-Zernike-like equation using a hypernetted chain (HNC) closure relation. Effective pair potentials are used to account for quantum effects. Results are compared with classical as well as quantum corrected Debye model calculations of Gregori~\textit{et~al.}~\cite{Gregori06b}. Especially, non-equilibrium situations with different electron and ion temperatures which are relevant on short time scales are studied motivated by the recent experiment~\cite{Glenzer06}.
\section{\label{theory}Theory}
The rigorous way to describe a multi-temperature plasma is to start from the BBGKY hierarchy~\cite{Bogoljubov46,Born46,Kirkwood35,Yvon35}. We are interested in the two-particle distribution function which allows, in principle, to calculate most of the physical quantities, e.g. pressure or internal energy. Here, we are especially interested in the structure factor, in order to interprete the scattering results.\par We consider a homogenous and isotropic system where binary interactions $V_{ab}$ and the two-particle distribution function $F_{ab}$ depend on the distance $\mathbf{r}_{12}=\mathbf{r}_1-\mathbf{r}_2$ of the two particles only. In absence of external forces the equation of motion for the two-particle distribution function $F_{ab}$ reads
\begin{align}
\label{theory:eq.1}\respace{\left(\frac{\partial}{\partial t}+\frac{\partial V_{ab}\left(r_{12}\right)}{\partial \mathbf{r}_1}\frac{\partial}{\partial \mathbf{p}_1}+\frac{\partial V_{ab}\left(r_{12}\right)}{\partial \mathbf{r}_2}\frac{\partial}{\partial \mathbf{p}_2}-\frac{\mathbf{p}_1}{m_a}\frac{\partial}{\partial \mathbf{r}_1}-\frac{\mathbf{p}_2}{m_b}\frac{\partial}{\partial \mathbf{r}_2}\right)F_{ab}\left(r_{12},\mathbf{p}_1,\mathbf{p}_2,t\right)}\\&=-\sum\limits_cn_c\int d^3\mathbf{r}_3\,d^3\mathbf{p}_3\,\frac{\partial V_{ac}\left(r_{13}\right)}{\partial \mathbf{r}_1} \frac{\partial}{\partial \mathbf{p}_1}F_{abc}\left(\mathbf{r}_1,\mathbf{r}_2,\mathbf{r}_3,\mathbf{p}_1,\mathbf{p}_2,\mathbf{p}_3,t\right)\nonumber\\&\quad-\sum\limits_cn_c\int d^3\mathbf{r}_3\,d^3\mathbf{p}_3\,\frac{\partial V_{bc}\left(r_{23}\right)}{\partial \mathbf{r}_2}\frac{\partial}{\partial \mathbf{p}_2}F_{abc}\left(\mathbf{r}_1,\mathbf{r}_2,\mathbf{r}_3,\mathbf{p}_1,\mathbf{p}_2,\mathbf{p}_3,t\right)\mbox{,}\nonumber
\end{align}
where indices $a$, $b$, and $c$ denote the plasma species. The quantities $n_c$ are the particle densities, and $F_{abc}$ is the three-particle distribution function, which itself couples to higher-order functions. One has to find an approximation to truncate equation~\eqref{theory:eq.1}.\par
As mentioned before we are interested in a model of the plasma, where each species $c$ is described within a local equilibrium with temperature $T_c$. This is possible if the momentum relaxation for particles of the same species is faster than for particles of different species. This fact is, for example, fulfilled having mass ratios far away from one between different species. Then the momentum distribution separates into a product of independent Maxwell distributions with temperature $T_c$. For the two-particle distribution function $F_{ab}$ one gets
\begin{equation}
\label{theory:eq.2}F_{ab}\left(r_{12},\mathbf{p}_1,\mathbf{p}_2\right)=g_{ab}\left(r_{12}\right)F_a\left(\mathbf{p}_1\right)F_b\left(\mathbf{p}_2\right)
\end{equation}
with $F_a\left(\mathbf{p}_1\right)=\exp{\left(-\beta_ap_1^2/\left(2m_a\right)\right)}/\left(2\pi m_a/\beta_a\right)^{3/2}$, where $m_c$ is the mass of the species $c$ and $\beta_c=\left(k_B T_c\right)^{-1}$ the inverse temperature. Higher-order distribution functions are defined likewise. Using this ansatz in equation~\eqref{theory:eq.1} yields
\begin{align}
\label{theory:eq.3}\respace{\left(\beta_a\frac{\mathbf{p}_1}{m_a}\frac{\partial V_{ab}\left(r_{12}\right)}{\partial \mathbf{r}_1}+\beta_b\frac{\mathbf{p}_2}{m_b}\frac{\partial V_{ab}\left(r_{12}\right)}{\partial \mathbf{r}_2}+\frac{\mathbf{p}_1}{m_a}\frac{\partial}{\partial \mathbf{r}_1}+\frac{\mathbf{p}_2}{m_b}\frac{\partial}{\partial \mathbf{r}_2}\right)g_{ab}\left(r_{12}\right)F_a\left(\mathbf{p}_1\right)F_b\left(\mathbf{p}_2\right)}\\&=-\beta_a\frac{\mathbf{p}_1}{m_a}\sum\limits_cn_c\int d^3\mathbf{r}_3\,\frac{\partial V_{ac}\left(r_{13}\right)}{\partial \mathbf{r}_1}g_{abc}\left(\mathbf{r}_1,\mathbf{r}_2,\mathbf{r}_3\right)F_a\left(\mathbf{p}_1\right)F_b\left(\mathbf{p}_2\right)\nonumber\\&\quad-\beta_b\frac{\mathbf{p}_2}{m_b}\sum\limits_cn_c\int d^3\mathbf{r}_3\,\frac{\partial V_{bc}\left(r_{23}\right)}{\partial \mathbf{r}_2}g_{abc}\left(\mathbf{r}_1,\mathbf{r}_2,\mathbf{r}_3\right)F_a\left(\mathbf{p}_1\right)F_b\left(\mathbf{p}_2\right)\mbox{.}\nonumber
\end{align}\par
Consequently, we concentrate on the two-particle spatial distribution function $g_{ab}$ only. We have first, in equation~\eqref{theory:eq.3}, to integrate over the momenta. Therefore equation~\eqref{theory:eq.3} is multiplied by $\mathbf{p}_1/m_a$ and by $\mathbf{p}_2/m_b$, respectively, yielding two equations. Performing the integration over the momenta for each equation results in
\begin{subequations}
\label{theory:eq.4}\begin{align}
\label{theory:eq.4a}\respace{\left(\frac{1}{m_a}\frac{\partial V_{ab}\left(r_{12}\right)}{\partial \mathbf{r}_{12}}+\frac{k_BT_a}{m_a}\frac{\partial}{\partial \mathbf{r}_{12}}\right)g_{ab}\left(r_{12}\right)}\\&=-\frac{1}{m_a}\sum\limits_cn_c\int d^3\mathbf{r}_3\,\frac{\partial V_{ac}\left(r_{13}\right)}{\partial \mathbf{r}_1}g_{abc}\left(\mathbf{r}_1,\mathbf{r}_2,\mathbf{r}_3\right)\nonumber\\\label{theory:eq.4b}\respace{\left(\frac{1}{m_b}\frac{\partial V_{ab}\left(r_{12}\right)}{\partial \mathbf{r}_{12}}+\frac{k_BT_b}{m_b}\frac{\partial}{\partial \mathbf{r}_{12}}\right)g_{ab}\left(r_{12}\right)}\\&=+\frac{1}{m_b}\sum\limits_cn_c\int d^3\mathbf{r}_3\,\frac{\partial V_{bc}\left(r_{23}\right)}{\partial \mathbf{r}_2}g_{abc}\left(\mathbf{r}_1,\mathbf{r}_2,\mathbf{r}_3\right)\mbox{.}\nonumber
\end{align}
\end{subequations}
Summing up equations~\eqref{theory:eq.4a} and~\eqref{theory:eq.4b} leads to
\begin{align}
\label{theory:eq.5}\respace{\frac{\partial h_{ab}\left(r_{12}\right)}{\partial \mathbf{r}_{12}}+\beta_{ab}\frac{\partial V_{ab}\left(r_{12}\right)}{\partial \mathbf{r}_{12}}g_{ab}\left(r_{12}\right)}\\&=-\beta_{ab}\frac{m_b}{m_a+m_b}\sum\limits_cn_c\int d^3\mathbf{r}_3\,\frac{\partial V_{ac}\left(r_{13}\right)}{\partial \mathbf{r}_1}g_{abc}\left(\mathbf{r}_1,\mathbf{r}_2,\mathbf{r}_3\right)\nonumber\\&\quad+\beta_{ab}\frac{m_a}{m_a+m_b}\sum\limits_cn_c\int d^3\mathbf{r}_3\,\frac{\partial V_{bc}\left(r_{23}\right)}{\partial \mathbf{r}_2}g_{abc}\left(\mathbf{r}_1,\mathbf{r}_2,\mathbf{r}_3\right)\nonumber
\end{align}
with $\beta_{ab}=(k_BT_{ab})^{-1}$ being the mass-weighted inverse temperature of two species $a$ and $b$ with $T_{ab}=(m_aT_b+m_bT_a)/(m_a+m_b)$ and $h_{ab}\equiv g_{ab}-1$ the total correlation function.\par
Result~\eqref{theory:eq.5} represents the equation of motion for the two-particle spatial distribution function $g_{ab}$. As noted earlier it is coupled to higher-order contributions in the hierarchy via the three-particle spatial distribution function $g_{abc}$. In order to solve equation~\eqref{theory:eq.5} a truncation is needed.\par
Assuming thermal equilibrium at the temperature $T$ equation~\eqref{theory:eq.5} reads
\begin{align}
\label{theory:eq.6}\respace{\frac{\partial h_{ab}\left(r_{12}\right)}{\partial \mathbf{r}_{12}}+\beta\frac{\partial V_{ab}\left(r_{12}\right)}{\partial \mathbf{r}_{12}}g_{ab}\left(r_{12}\right)}\\&=-\frac{m_b}{m_a+m_b}\sum\limits_cn_c\int d^3\mathbf{r}_3\,\beta\frac{\partial V_{ac}\left(r_{13}\right)}{\partial \mathbf{r}_1}g_{abc}\left(\mathbf{r}_1,\mathbf{r}_2,\mathbf{r}_3\right)\nonumber\\&\quad+\frac{m_a}{m_a+m_b}\sum\limits_cn_c\int d^3\mathbf{r}_3\,\beta\frac{\partial V_{bc}\left(r_{23}\right)}{\partial \mathbf{r}_2}g_{abc}\left(\mathbf{r}_1,\mathbf{r}_2,\mathbf{r}_3\right)\mbox{.}\nonumber
\end{align}
It is known that equation~\eqref{theory:eq.6} can be decoupled formally by introduction of the direct correlation function $c_{ab}$, giving
\begin{align}
\label{theory:eq.7}\respace{\frac{\partial h_{ab}\left(r_{12}\right)}{\partial \mathbf{r}_{12}}-\frac{\partial c_{ab}\left(r_{12}\right)}{\partial \mathbf{r}_{12}}}&=\frac{m_b}{m_a+m_b}\sum\limits_cn_c\int d^3\mathbf{r}_3\,\frac{\partial c_{ac}\left(r_{13}\right)}{\partial \mathbf{r}_1}h_{cb}\left(r_{23}\right)\\&\quad-\frac{m_a}{m_a+m_b}\sum\limits_cn_c\int d^3\mathbf{r}_3\,h_{ca}\left(r_{13}\right)\frac{\partial c_{bc}\left(r_{23}\right)}{\partial \mathbf{r}_2}\mbox{,}\nonumber
\end{align}
which is equivalent to the Ornstein-Zernike equation~\cite{Ornstein14}
\begin{equation}
\label{theory:eq.8}h_{ab}\left(r_{12}\right)-c_{ab}\left(r_{12}\right)=\sum\limits_cn_c\int d^3\mathbf{r}_3\,c_{ac}\left(r_{13}\right)h_{cb}\left(r_{23}\right)\mbox{.}
\end{equation}
To lowest order in the interaction, one has $c_{ab}^0\approx h_{ab}^0\approx-\beta V_{ab}$.\par
For the multi-temperature regime, it is reasonable that decoupling the hierarchy by a direct correlation function $c_{ab}$ fulfills, to lowest order, the relation $c_{ab}^0\approx h_{ab}^0\approx-\beta_{ab}V_{ab}$. Thus the corresponding multi-temperature Ornstein-Zernike equation reads
\begin{align}
\label{theory:eq.9}\respace{\frac{\partial h_{ab}\left(r_{12}\right)}{\partial \mathbf{r}_{12}}-\frac{\partial c_{ab}\left(r_{12}\right)}{\partial \mathbf{r}_{12}}}&=+\beta_{ab}\frac{m_b}{m_a+m_b}\sum\limits_cn_c\int d^3\mathbf{r}_3\,\frac{1}{\beta_{ac}}\frac{\partial c_{ac}\left(r_{13}\right)}{\partial \mathbf{r}_1}h_{cb}\left(r_{23}\right)\\&\quad-\beta_{ab}\frac{m_a}{m_a+m_b}\sum\limits_cn_c\int d^3\mathbf{r}_3\,\frac{1}{\beta_{bc}}\frac{\partial c_{bc}\left(r_{23}\right)}{\partial \mathbf{r}_2}h_{ca}\left(r_{23}\right)\mbox{.}\nonumber
\end{align}
In order to solve equation~\eqref{theory:eq.9} a closure relation is needed, connecting the direct correlation function $c_{ab}$ to the total one $h_{ab}$. In equilibrium there exist various closure relations~\cite{Percus58, Lebowitz64, Leeuwen59, Morita60,Lebowitz66} that can be easily transferred to a multi-temperature plasma by the same way described above, i.e. introducing the mass-weighted temperature $T_{ab}$ for the species $a$ and $b$. We use a HNC-like closure relation written as
\begin{equation}
\label{theory:eq.10}c_{ab}\left(r_{12}\right)=h_{ab}\left(r_{12}\right)-\ln{g_{ab}\left(r_{12}\right)}-\beta_{ab}V_{ab}\left(r_{12}\right)\mbox{.}
\end{equation}\par
The non-linear set of equations~\eqref{theory:eq.9} and~\eqref{theory:eq.10} is solved in the usual way~\cite{Springer73} giving results for the direct and total correlation functions, $c_{ab}$ and $h_{ab}$. The structure factor $S_{ab}$ is defined by the well-known relation
\begin{align}
\label{theory:eq.11}S_{ab}\left(k\right)&=\delta_{ab}+\sqrt{n_an_b}\int d^3\mathbf{r}_{12}\exp{\left(-i\mathbf{k}\mathbf{r}_{12}\right)}\left(g_{ab}\left(r_{12}\right)-1\right)\\&=\delta_{ab}+\sqrt{n_an_b}\,\tilde{h}_{ab}\left(k\right)\mbox{.}\nonumber
\end{align}
\section{\label{resultsanddiscussion}Results and Discussion}
We have considered a two-component beryllium plasma consisting of electrons and ions with an effective charge state Z. For the description of the interactions $V_{ab}$ there exist numerous effective pair potentials, e.g. see~\cite{Kelbg63a,Kelbg63b,Deutsch77,Filinov04}. For all these potentials, the main goal is to incorporate short-range interactions to avoid divergencies which result from the behaviour of the Coulomb potential $V_{ab}\left(r_{12}\right)=\left(q_aq_b\right)/\left(4\pi\varepsilon_0r_{12}\right)$ at zero distance. In classical schemes, this problem may be solved by hard-sphere-like potentials. An improvement is given by using pseudopotential theory as known from solid state physics. In a first simple step we apply quantum mechanics by evaluating the two-particle Slater sum $\mathbb{S}_{ab}$. One gets an effective potential $V_{ab}$, see Morita~\cite{Morita59}, via
\begin{equation}
\label{resultsanddiscussion:eq.1}\mathbb{S}_{ab}\left(r_{12}\right)=\exp{\left(-\beta_{ab}V_{ab}\left(r_{12}\right)\right)}\mbox{.}
\end{equation}
The two-particle slater sum itself is defined as
\begin{equation}
\label{resultsanddiscussion:eq.2}\mathbb{S}_{ab}\left(\mathbf{r}_{12}\right)=8\pi^\frac{3}{2}\lambda_{ab}^3\left[1+\frac{\left(-1\right)^{2s_a}\delta_{ab}}{2s_a+1}\right]\sum\limits_\alpha\exp{\left(-\beta_{ab}E_\alpha\right)}\left|\Psi_\alpha \left(r_{12},\theta,\phi\right)\right|^2\mbox{.}
\end{equation}
Here the thermal wavelengths $\lambda_{ab}=\hbar\sqrt{\beta_{ab}/\left(2 m_{ab}\right)}$ with $m_{ab}^{-1}=m_a^{-1}+m_b^{-1}$ and the spin quantum number $s_a$ are introduced. The quantity $\alpha$ stands for the quantum numbers $n$,$l$, and $m$ for bound states and $k$,$l$, and $m$ for scattering states with $\Psi_\alpha$ being the corresponding wave function, respectively.\par
The first-order Born approximation at low degeneracy in equation~\eqref{resultsanddiscussion:eq.2} leads to the Kelbg potential~\cite{Kelbg63a}
\begin{equation}
\label{resultsanddiscussion:eq.3}
V_{ab}\left(r_{12}\right)=\frac{q_aq_b}{4\pi\varepsilon_0r_{12}}\left[1-\exp{\left(-\frac{r_{12}^2}{\lambda_{ab}^2}\right)}+\sqrt{\pi}\frac{r_{12}}{\lambda_{ab}}\erfc{\left(\frac{r_{12}}{\lambda_{ab}}\right)}\right]
\end{equation}
with $\erfc{\left(x\right)}$ being the standard complementary error function. This type of potential has already been considered in \cite{Filinov04}.\par
For the electron-ion interaction one has to exclude the effect of bound states. We applied in the subsequent calculations a potential derived by Klimontovich and Kraeft~\cite{Klimontovich74} (a simplified version of their equation (42))
\begin{equation}
\label{resultsanddiscussion:eq.4}
V_{ei}\left(r_{12}\right)=-\frac{k_B T_{ei}\xi_{ei}^2}{16}\left(1+\frac{k_B T_{ei}\xi_{ei}^2}{16\frac{q_eq_i}{4\pi\varepsilon_0}}r\right)^{-1}\mbox{,}
\end{equation}
where $\xi_{ei}=\left(q_eq_i\beta_{ei}\right)/\left(4\pi\varepsilon_0\lambda_{ei}\right)$. In this potential, only the ground state of a bound electron is excluded. Both the Kelbg potential~\eqref{resultsanddiscussion:eq.3} and the Klimontovich-Kraeft potential~\eqref{resultsanddiscussion:eq.4} are finite at zero distance. The equations~\eqref{resultsanddiscussion:eq.1},~\eqref{resultsanddiscussion:eq.2},~\eqref{resultsanddiscussion:eq.3}, and~\eqref{resultsanddiscussion:eq.4}, derived for thermal equilibrium, are generalised to a two-temperature situation by replacing the temperature $T$ with the mass-weighted temperature $T_{ab}$ as explained in section~\ref{theory}. The interaction potentials $V_{ab}$ are displayed in figure~\ref{resultsanddiscussion:fig.1}.\par
\begin{vchfigure}[htb]
\centerline{\includegraphics[width=0.5\textwidth]{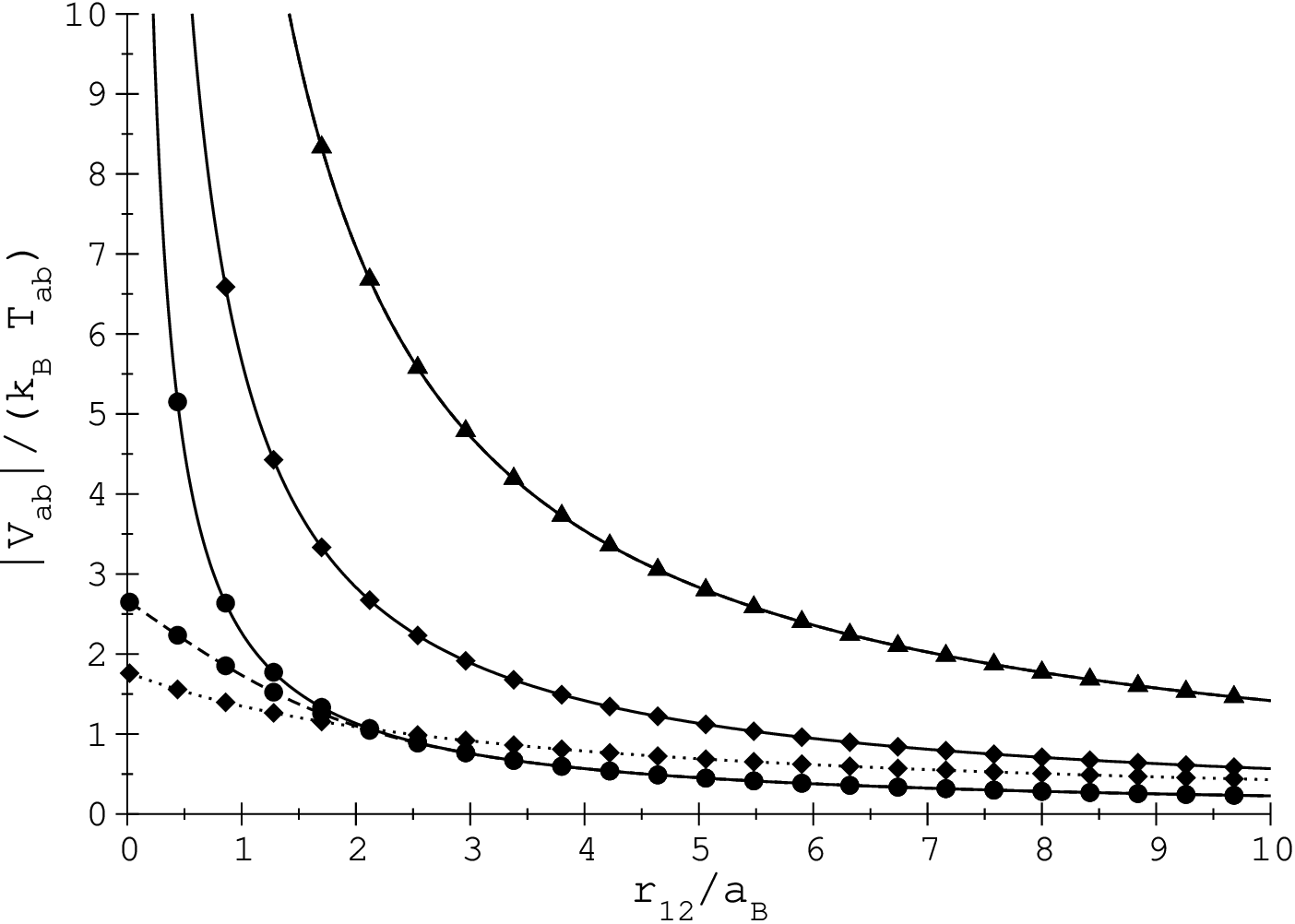}}
\vchcaption{\label{resultsanddiscussion:fig.1}Effective pair potentials $\left|V_{ab}\right|$ used in the calculations for a beryllium plasma shown against the distance $r_{12}$ for charge state $Z=2.5$ and fixed electron and ion temperature at $12$~eV. Notations are as follows: solid line for the Coulomb potential, dashed line for the Kelbg potential~\eqref{resultsanddiscussion:eq.3}, and dotted line for the Klimontovich-Kraeft potential~\eqref{resultsanddiscussion:eq.4}. The interactions are marked by a filled circle for the electron-electron, diamond for the electron-ion, and triangle for the ion-ion interaction. For the ion-ion interaction the Kelbg potential is close to the Coulomb potential, but has a finite value at zero distance.}
\end{vchfigure}
We calculated the ionic structure factor $S_{ii}$ within the formalism derived in section~\ref{theory}. For the description of the electron-electron and ion-ion interaction we used the Kelbg potential~\eqref{resultsanddiscussion:eq.3}. Because we found the Kelbg potential not applicable for the electron-ion interaction in that low temperature region, we applied the Klimontovich-Kraeft potential~\eqref{resultsanddiscussion:eq.4}. Results for the ionic structur factor $S_{ii}$ are shown in figure~\ref{resultsanddiscussion:fig.2} versus the wave number $k$ for a beryllium plasma with charge state $Z=2.5$, ion density $n_i=1.21\times10^{23}$~cm$^{-3}$, and fixed electron temperature $T_e=12$~eV. We compare calculations for different ratios of electron $T_e$ and ion $T_i$ temperature. Also shown are results by Gregori~\textit{et~al.}~\cite{Gregori06b} and for the Debye approximation for a pure ionic system, $S_{ii}\left(k\right)=k^2/\left(k^2+\kappa_{ii}^2\right)$ with $\kappa_{ii}^2=\left(Z^2\mbox{e}^2n_i\right)/\left(\varepsilon_0k_BT_i\right)$, which can be considered as the limit for weak correlations.\par
\begin{vchfigure}[htb]
\centerline{\includegraphics[width=0.5\textwidth]{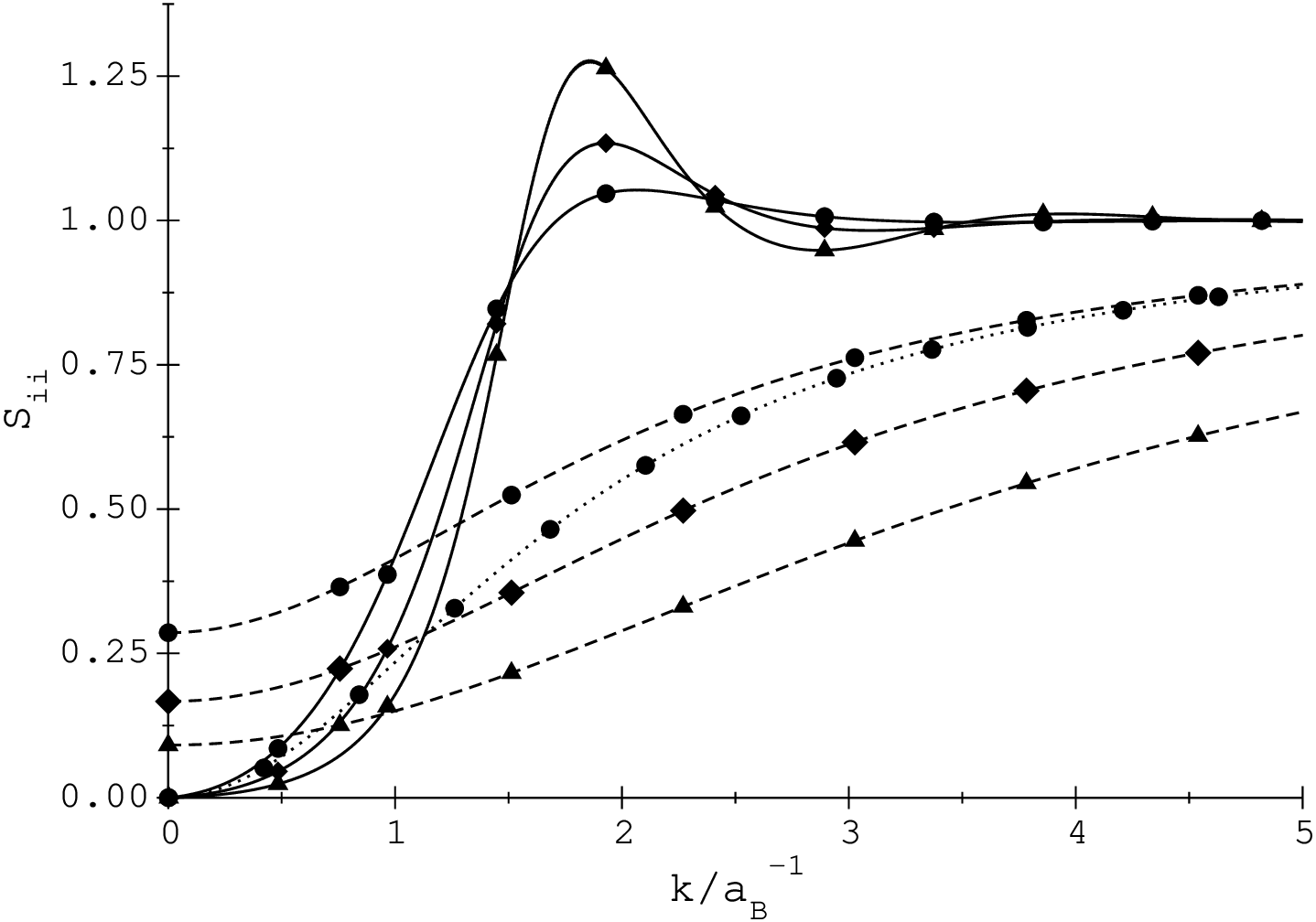}}
\vchcaption{\label{resultsanddiscussion:fig.2}Ion-ion structure factor $S_{ii}$ for a beryllium plasma with charge state $Z=2.5$, ion density $n_i=1.21\times10^{23}$~cm$^{-3}$, and fixed electron temperature $T_e= 12$~eV plotted versus the wave number $k$. The result of the present work is displayed as solid line. The work of Gregori~\textit{et~al.}~\cite{Gregori06b} and the Debye result are shown with dashed and dotted line, respectively. Three different temperature ratios $T_e/T_i$ between electron $T_e$ and ion $T_i$ temperature were examined marked  as filled circle for $T_e/T_i=1$, diamond for $T_e/T_i=2$, and triangle for $T_e/T_i=4$. The Debye approximation is plotted for $T_i=12$~eV.}
\end{vchfigure}
In contrast to the present paper, Gregori~\textit{et~al.}~\cite{Gregori06b} did not use the HNC equations, but generalised the screened pseudopotential approach of Arkhipov and Davletov~\cite{Arkhipov98} for a two-temperature system. In this approximation the ionic structure factor is given by $S_{ii}\left(k\right)=1-n_i\tilde{\Phi}_{ii}\left(k\right)/\left(k_B T_i^\prime\right)$. The effective ion temperature $T_i^\prime=\sqrt{T_i^2+0.152\,T_D^2}$ with $T_D$ being the Debye temperature accounts for ion degeneracy at low temperatures. The pseudopotential $\Phi_{ii}$ is a modified Kelbg-Deutsch potential~\cite{Kelbg63b,Deutsch77} with Debye-like screening corrections.\par
Consider in figure~\ref{resultsanddiscussion:fig.2} the curves for the temperature ratio $T_e/T_i=1$: For small $k$, the present result is close to the Debye result. Our calculations show a small but nevertheless non zero value at $k=0$. With increasing $k$, the ionic structure factor is increasing rapidly showing a peak at $k\approx1.8\,a_B^{-1}$. For values of $k$ greater than $k\approx4.5\,a_B^{-1}$, the function is almost one, that means the system is uncorrelated at that scale. The result of Gregori~\textit{et~al.} clearly shows a different behaviour. Their curve starts with a finite value around $0.3$ at $k=0$ indicating long range polarisation effects. With increasing $k$ the function increases slower showing no peak and turning into a Debye-like behaviour.\par
Concerning the temperature dependence there is a common trend in the results of the present work and those of Gregori~\textit{et~al.}~\cite{Gregori06b}. With decreasing ion temperature $T_i$, the value of the ionic structure factor $S_{ii}$ for small $k$ is lowered. This can be attributed to a weakening of the correlations between the electrons and ions. The system is approaching towards a one-component plasma (OCP) with static ions in a screening background. Within the HNC calculations this leads to a sharper and stronger peak. The occurrence of such structures in our calculations is a progress in the description of multi-temperature plasma resulting from the inclusion of higher-order correlations, see for comparison~\cite{Springer73}.\par
\begin{vchfigure}[htb]
\centerline{\includegraphics[width=0.5\textwidth]{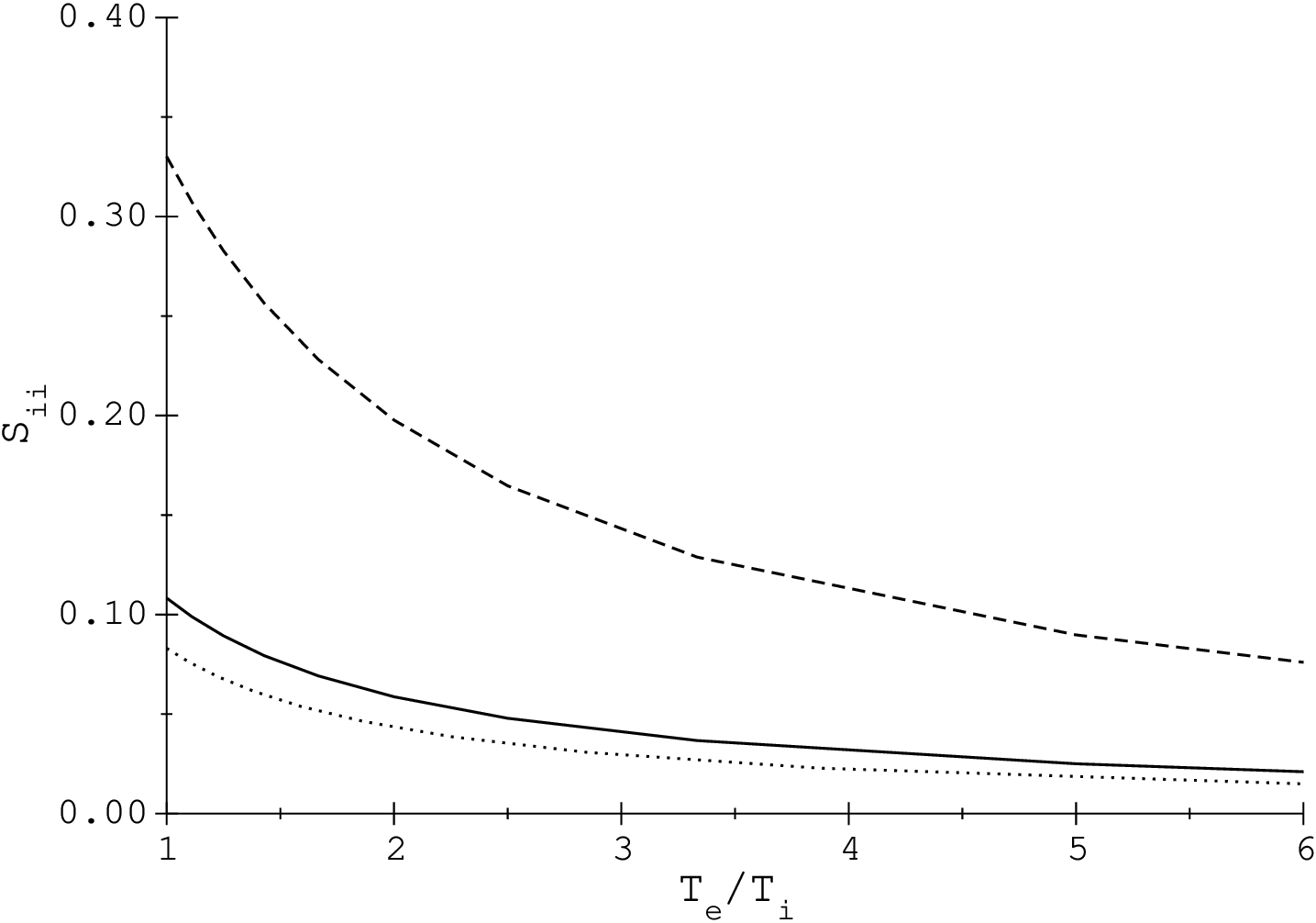}}
\vchcaption{\label{resultsanddiscussion:fig.3}Ion-ion structure factor $S_{ii}$ for a beryllium plasma as a function of the ion temperature $T_i$. The electron temperature $T_e$ is fixed at $12$~eV, the ion density is $n_i=1.21\times10^{23}$~cm$^{-3}$, charge state is $Z=2.5$, and $k=0.54\,a_B^{-1}$. The result of the present work is displayed as solid line. The work of Gregori~\textit{et~al.}~\cite{Gregori06b} and the Debye result are shown with dashed and dotted lines, respectively.}
\end{vchfigure}
For the specific experiment~\cite{Glenzer06}, the ionic structure factor at $k=0.54\,a_B^{-1}$ is of interest. For Thomson scattering at a wavelength of $\lambda=0.42$~nm, the $k$ value corresponds to a scattering angle of $\theta=40^o$, i.e. forward scattering. For this $k$ value the ionic structure factor $S_{ii}$ is considered as function of the ion temperature $T_i$ and of the charge state Z in figures~\ref{resultsanddiscussion:fig.3} and~\ref{resultsanddiscussion:fig.4}, respectively. Figure~\ref{resultsanddiscussion:fig.3} shows, for this small $k$ value, a systematic decrease of the structure factor for decreasing ion temperature. The absolute values however are model dependent.\par
The dependence on the charge state is plotted in figure~\ref{resultsanddiscussion:fig.4} for $T_e=T_i=12$~eV. The curves look similar to figure~\ref{resultsanddiscussion:fig.3}. This can be understood easily keeping in mind that a lower ion temperature as well as an increased charge number Z are connected with a stronger coupling of the ions in the system.
\begin{vchfigure}[htb]
\centerline{\includegraphics[width=0.5\textwidth]{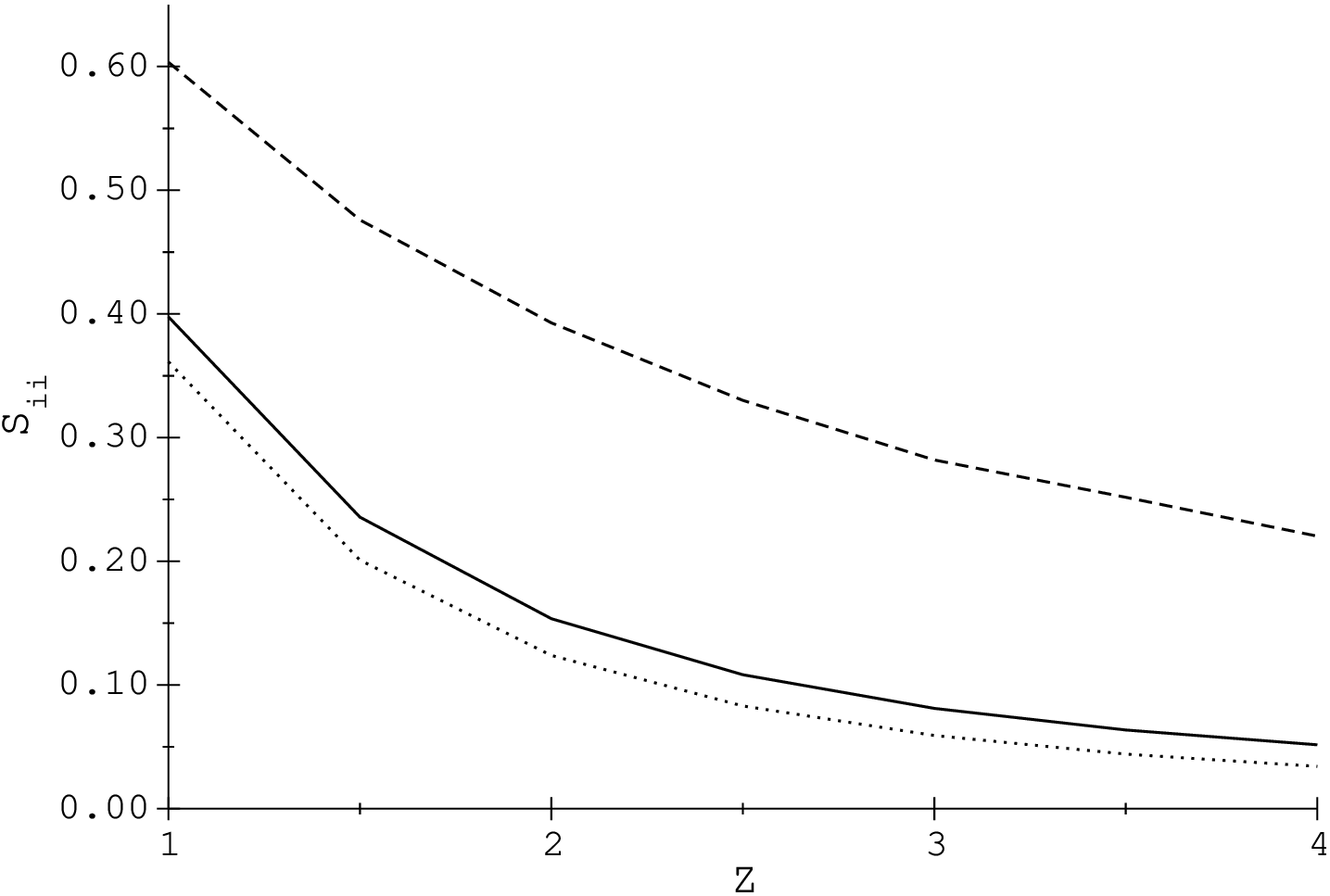}}
\vchcaption{\label{resultsanddiscussion:fig.4}Ion-ion structure factor $S_{ii}$ for a beryllium plasma as a function of the ion charge state Z. The electron temperature $T_e$ as well as the ion temperature $T_i$ are $12$~eV, the ion density is $n_i=1.21\times10^{23}$~cm$^{-3}$, and $k=0.54\,a_B^{-1}$. The result of the present work is displayed as solid line. The work of Gregori~\textit{et~al.}~\cite{Gregori06b} and the Debye result are shown with dashed and dotted lines, respectively.}
\end{vchfigure}
\section{Conclusions}
We have calculated the static structure factor $S\left(k\right)$ for solid-density beryllium consisting of electrons and ions. Effective pair potentials were used to include quantum mechanical corrections. We studied non-equilibrium situations for a two-temperature plasma. In this case a mass weighted temperature occurs. We find a strong influence of correlations on the static structure factor compared with simpler approaches such as the Debye model with quantum mechanical corrections of Gregori~\textit{et~al.}~\cite{Gregori06b} and the classical Debye model. Therefore, the correct treatment of all correlations in a two-component plasma is crucial for the description of non-equilibrium states and their relaxation into equilibrium. A consistent derivation especially of the effective electron-ion and ion-ion potentials, is, of course, needed and remains the subject of further work, e.g. accounting for finite size corrections.
\begin{acknowledgement}
We thank G.~Gregori and D.~Kremp for their helpful discussions. This work is supported by the DFG SFB 652 "Strong correlations and collective effects in radiation fields: Coulomb systems, clusters, and particles", and by the Helmholtz-Gemeinschaft Virtual Institute VH-VI-104 "Plasma Physics Using FEL Radiation". The work of S. H. Glenzer was performed under the auspices of the U.~S.~Department of Energy by the University of California, Lawrence Livermore National Laboratory under contract No.~W-7405-Eng-48. S. H. Glenzer was also supported by 05-LDRD-003 and the Alexander von Humboldt Foundation.
\end{acknowledgement}

\bibliography{cpp-paper}
\bibliographystyle{prsty}

\end{document}